\documentclass[prd,eqsecnum]{revtex4}

\newcommand{\beq}{\begin{equation}}
\newcommand{\eeq}{\end{equation}}
\newcommand{\beqa}{\begin{eqnarray}}
\newcommand{\eeqa}{\end{eqnarray}}
\newcommand{\lslash}[1]{#1\llap/}
\newcommand{\Eq}[1]{Eq.\ (\ref{#1})}
\newcommand{\Eqs}[2]{Eqs.\ (\ref{#1}) and (\ref{#2})}
\newcommand{\Ref}[1]{Ref.\ \cite{#1}}

\begin{document}

\title{Electromagnetic annihilation into charged leptons and scattering
off nucleons of spin-3/2 Majorana particles}

\author{Jos\'e F. Nieves}
\email{nieves@ltp.uprrp.edu}
\affiliation{Laboratory of Theoretical Physics\\
Department of Physics, University of Puerto Rico \\
R\'{\i}o Piedras, Puerto Rico 00936}
\author{Sarira Sahu}
\email{sarira@nucleares.unam.mx}
\affiliation{Instituto de Ciencias Nucleares, Universidad Nacional
  Autonoma de Mexico,\\
Circuito Exterior, C. U., A. Postal 70-543, 04510 Mexico DF, Mexico}

\date{July 2014}

\begin{abstract}
We compute the cross section for the electromagnetic annihilation
into charged leptons, and the electromagnetic scattering off nucleons,
of spin-3/2 self-conjugate (Majorana) particles using the general form of the
electromagnetic vertex function that was obtained previously for
such particles. In addition to the restrictions imposed  by common
principles such as electromagnetic gauge invariance and hermiticity,
the vertex function incorporates the restriction due to the
Majorana condition as well as the particular properties related to the
spinors in the Rarita-Schwinger representation, and is the counterpart
of the so-called \emph{anapole} interaction of spin-1/2 Majorana particles.
The formulas obtained for the cross sections share certain similarities
with the corresponding results in the spin-1/2 case,
but they also reveal some important differences which are pointed out
and discussed. The results given here can be useful for applications
involving the electromagnetic interactions of spin-3/2 or spin-1/2
Majorana particles in several contexts that have been of interest
in the recent literature such as nucleosynthesis and dark matter.
\end{abstract}
\maketitle

\section{Introduction and Summary}

The study of the electromagnetic processes of electrically neutral particles
has been of interest in a variety of physical and astrophysical contexts.
This is largely due to the fact that since they do not couple
to the photon at the tree-level, their electromagnetic properties
can provide a window to higher order effects or sectors of
the Standard Model (SM) and its extensions that may be difficult
or impossible to probe directly. This is even more so in the
particular case of Majorana particles. 
For example, it has been known for a long time that Majorana neutrinos
can have neither electric nor magnetic dipole moments; it can only
have the so-called axial charge radius. Results of this type can also
be deduced, for example, for the transition moments between two
different Majorana neutrinos\cite{n:nuem,k:nuem,s:nuem,kg:nuem,k:nucpt},
and similar results hold for spin-1 particles\cite{np:spin1}
and for Majorana particles of arbitrary spin\cite{boudjemahanzaoui}.
By the same token, the experimental observation of departures
from these results would have implications for some important principles
such as Lorentz and gauge invariance, $CPT$ and crossing symmetry.

Motivated by this, we recently considered in a general way the electromagnetic
properties of spin-3/2 Majorana particles\cite{n:spin32m},
to provide a systematic study that while being helpful for
considering questions of fundamental
and intrinsic interest, is also useful for phenomenological
applications in several areas of current interest.
For example, there has been considerable activity recently in the study of
the possible effects of gravitinos in several cosmological
contexts such as nucleosynthesis and
inflation\cite{inflation1,inflation2,cosmology1,nucleosynthesis1}, and
the implications of spin-3/2 particles as dark matter
candidates and their detection\cite{darkmatter1,darkmatter2,darkmatter3}.
Many of these effects have to do with the electromagnetic properties
of gravitinos, or more generally with the electromagnetic
properties of spin-3/2 Majorana particles.

In the present work we calculate the cross section
for the electromagnetic annihilation into charged leptons,
and the electromagnetic scattering off nucleons,
by a spin-3/2 self-conjugate (Majorana) particle
using the results obtained in \Ref{n:spin32m} for the electromagnetic
vertex function of such a particle.
The analogous results for spin-1/2 Majorana particles, which involve
the so-called \emph{anapole} interaction,
have been used recently in detailed analysis in
Refs.\ \cite{scherrer1,scherrer2} to constrain their properties.
As we show, the cross sections in the spin-3/2 case share certain similarities
with the corresponding results in the spin-1/2 case,
but they also reveal some important differences.
The results presented here can be useful in the contexts already
mentioned above, and they can be also used to extend to the spin-3/2
case the work carried out in Refs.\ \cite{scherrer1,scherrer2}
for spin-1/2.

To be more precise, let us summarize the
results obtained in \Ref{n:spin32m} that are most relevant for the present
work. Borrowing from the notation introduced there,
we denote by $j_\mu(Q,q)$ the matrix element of the electromagnetic
current operator $J^{\rm (em)}_\mu(0)$ between two spin-3/2 particle states,
with momentum $k,k^\prime$ and spin projection $\sigma,\sigma^\prime$,
\begin{eqnarray}
\label{defmatrixelement}
j_\mu(Q,q) \equiv
\langle k^\prime,\sigma^\prime|J^{\rm (em)}_\mu(0)|k,\sigma\rangle\,,
\end{eqnarray}
where
\begin{eqnarray}
q = k - k^\prime \,,\nonumber\\
Q = k + k^\prime\,.
\end{eqnarray}
The electromagnetic off-shell vertex function
$\Gamma_{\alpha\beta\mu}(Q,q)$ is then defined such that
\begin{eqnarray}
\label{jGammadef}
j_\mu(Q,q) = \bar U^{\alpha}(k^\prime,\sigma^\prime)
\Gamma_{\alpha\beta\mu}(Q,q)
U^\beta(k,\sigma) \,,
\end{eqnarray}
where $U^\alpha(k,\sigma)$ is a Rarita-Schwinger (RS) spinor.

Then, for the particular case in which the initial and final particle is the
same and it is self-conjugate (the \emph{diagonal Majorana case}),
which denote by $\lambda$, the most general form of the vertex function
consistent with Lorentz and electromagnetic gauge invariance,
involves at most two independent terms, which can be written in the form
\begin{eqnarray}
\label{majdiagvertex1}
\Gamma_{\alpha\beta\mu} & = & f
g_{\alpha\beta}\left(q^2\gamma_\mu - q_\mu\lslash{q}\right)\gamma_5
+ \frac{if}{m_\lambda}\left(q_\beta\epsilon_{\alpha\mu\nu\lambda} -
q_\alpha\epsilon_{\beta\mu\nu\lambda}\right)q^\nu Q^\lambda
\nonumber\\
&&\mbox{} + ig\left[
q^2 (g_{\mu\alpha} q_\beta + g_{\mu\beta} q_\alpha) -
2q_\mu q_\alpha q_\beta\right] \,,
\end{eqnarray}
where $m_\lambda$ is the $\lambda$ particle's mass
and the two corresponding form factors, denoted here by $f$ and $g$, are real.
The term proportional to $\gamma_\mu\gamma_5$ is reminiscent
of the axial charge radius term for Majorana
neutrinos\cite{n:nuem,k:nuem,s:nuem}, while the other two terms resemble the
result that was obtained in Ref.\ \cite{np:spin1} for the electromagnetic
vertex function of self-conjugate spin-1 particles [Eq. (4.14) in that
reference]. Further conditions exist if some discrete symmetries hold.
For example, if $CP$ holds, then $g = 0$, so that only the firt two terms
in \Eq{majdiagvertex1}, involving the $f$ form factor, can be present.
An expression similar to \Eq{majdiagvertex1}
was obtained in \Ref{boudjemahanzaoui}, where it was found that
the vertex function in this same case (spin-3/2 diagonal Majorana case)
consists of at most three terms.
However, the distinctive feature of the result obtained in \Ref{n:spin32m}
that we are emphasizing here, is the fact that the coefficients of the first
two terms in \Eq{majdiagvertex1} are not independent, so that at most
two form factors are involved. In particular,
if the $\gamma_\mu \gamma_5$ is present,
then it must be accompanied by the second term in
\Eq{majdiagvertex1}. This result is due to the requirement that
the vertex function does not mix the genuine spin-3/2 degrees of freedom
with the spurious spin-1/2 components of the RS representation.
Stated more precisely, if we denote by $(S_{\mu\nu})_{\alpha\beta}$ and
$\frac{1}{2}\sigma_{\mu\nu} g_{\alpha\beta}$ the spin-1 and spin-1/2
Lorentz group generators in the RS representation, the above result
follows by requiring that these generators do not appear in
the vertex function separately, but in the combination
$(\Sigma_{\mu\nu})_{\alpha\beta} = (S_{\mu\nu})_{\alpha\beta} +
\frac{1}{2}\sigma_{\mu\nu} g_{\alpha\beta}$, and its products.

In this work we use the expression given in \Eq{majdiagvertex1}
as the starting point to compute the cross section
for the annihilation process
\beq
\lambda + \lambda \rightarrow \ell + \bar \ell\,,
\eeq
where $\ell$ stands for a charged lepton (e.g., the electron),
and for the scattering process
\beq
\lambda + f \rightarrow \lambda + f
\eeq
where $f = \ell, n, p$ is a charged lepton or a nucleon.
We restrict ourselves to the case $g = 0$, that is we assume that
in these processes the $CP$ symmetry holds. We begin in
Section\ \ref{sec:preliminaries} with a summary of the notation
and conventions, and in particular with our conventions regarding
the spin-3/2 particle spinor wave functions. In
Section\ \ref{sec:annihilation} we carry out the calculation
of the annihilation cross section and present the formula
for the differential cross section and in Section\ \ref{sec:scattering}
we present the corresponding calculations and results for the scattering
process. Because the calculational algebra that involves using the RS
spinors can be cumbersome and lengthy, some details of the calculations
in those two sections are provided in three appendices.
In Section\ \ref{sec:conclusions} we discuss
the salient features of the results we have obtained
and compare them with the cross sections for the analogous processes
involving a spin-1/2 Majorana or a
spin-1 self-conjugate particle, denoted by $\chi$ and $V$ respectively.
We close that section with some concluding remarks and outlook regarding
the processes $\lambda + \lambda \rightarrow \gamma + \gamma$ and
$\lambda + \gamma \rightarrow \lambda + \gamma$. In contrast with
the analogous $\chi$ processes, the tree-level amplitudes
for both of these processes do not vanish and therefore
it could be of interest to consider them along similar lines.

\section{Notation and Preliminaries}
\label{sec:preliminaries}

\subsection{Kinematics}

\subsubsection{Scattering process}

To establish the notation and conventions let us consider the scattering
process
\begin{equation}
\lambda(k) + f(p) \rightarrow \lambda(k^\prime) + f(p^\prime)\,,
\end{equation}
where the momentum vectors have components
\beqa
\label{componentnotation}
k^\mu & = & (\omega,\vec k) \,,\nonumber\\
p^\mu & = & (E,\vec p) \,,
\eeqa
with similar notation for the primed counterparts,
satisfying the on-shell conditions
\beqa
\label{onshellk}
k^2 & = k^{\prime\,2} = m^2_\lambda\,,\nonumber\\
p^2 & = p^{\prime\,2} = m^2_f\,.
\eeqa

Denoting by $q$ the virtual photon momentum,
\beq
\label{defqscatt}
q = k - k^\prime = p^\prime - p\,,
\eeq
and defining
\beqa
\label{PQdefscatt}
Q & = & k + k^\prime\,,\nonumber\\
P & = & p + p^\prime \,,
\eeqa
the following relations hold,
\beqa
\label{Q2scatt}
\label{P2scatt}
k\cdot k^\prime & = & m^2_\lambda - \frac{1}{2}t\,,\nonumber\\
Q^2 & = & 4m^2_\lambda - t\,,\nonumber\\
p\cdot p^\prime & = & m^2_f - \frac{1}{2}t\,,\nonumber\\
P^2 & = & 4m^2_f - t\,,
\eeqa
where
\beq
\label{t}
t = q^2 \,.
\eeq

\subsubsection{Annihilation process}

In the annihilation channel,
\beq
\label{conventionsann}
\lambda(k) + \lambda(k^\prime) \rightarrow \ell(p) + \bar \ell(p^\prime)\,.
\eeq
the virtual photon momentum is
\beq
\label{qdefann}
q = k + k^\prime = p + p^\prime \,,
\eeq
and
\beq
\label{Qdefann}
Q = k - k^\prime\,.
\eeq
Assuming again the on-shell conditions, the following relations then hold
\beqa
\label{kcdotkprimeann}
k\cdot k^\prime & = & \frac{1}{2}s - m^2_\lambda\,,\\
\label{Q2ann}
Q^2 & = & 4m^2_\lambda - s\,,
\eeqa
where
\beq
\label{defs}
s = q^2 \,.
\eeq

\subsection{Spinor conventions and relations}

The spin-1/2 Dirac spinors are denoted by $u,v$,
while $U,V$ denote the spin-3/2 Rarita-Schwinger spinors.
The Dirac spinors are normalized in the usual
way so that in particular the polarization sums are given by the standard
formulas
\beqa
\label{spin12polsum}
\rho(p) \equiv \sum_s u(p,s)\,\bar u(p,s) & = & \lslash{p} + m_f\,,
\nonumber\\
\bar\rho(p) \equiv \sum_s v(p,s)\,\bar v(p,s) & = & \lslash{p} - m_f\,.
\eeqa
The spin-3/2 spinors are normalized such that
\beq
-\bar V^\mu V_\mu = \bar U^\mu U_\mu = -2m_{\lambda} \,,
\eeq
where the minus sign in the right-hand-side is analogous
to the minus sign in the normalization of the spin-1
polarization vectors, $\epsilon^\mu \epsilon_{\mu} = -1$.
The formulas for the polarization sums corresponding to \Eq{spin12polsum} are
\beqa
\label{spin32poloper}
\rho_{\mu\nu}(k) & \equiv & \sum_\sigma U_\mu(k,\sigma) \bar U_\nu(k,\sigma) =
-(\lslash{k} + m_\lambda)R_{\mu\nu}(k) \,,\nonumber\\
\bar \rho_{\mu\nu}(k) & \equiv &
\sum_\sigma V_\mu(k,\sigma) \bar V_\nu(k,\sigma) =
-(\lslash{k} - m_\lambda)R_{\mu\nu}(k) \,,
\eeqa
where
\beq
\label{defR}
R_{\mu\nu}(k) = \tilde g_{\mu\nu}(k) -
\frac{1}{3}\tilde\gamma_\mu(k) \tilde\gamma_\nu(k) \,,
\eeq
with
\beq
\label{gtilde}
\tilde g_{\mu\nu}(k) = g_{\mu\nu} - \frac{k_\mu k_\nu}{k^2}
\eeq
and
\beq
\label{gammatilde}
\tilde\gamma_\mu(k) = \tilde g_{\mu\alpha}(k) \gamma^\alpha \,.
\eeq
The tensor $R_{\mu\nu}(k)$ is transverse to $k$, and also satisfies
the RS auxiliary conditions
\beqa
\gamma^\mu R_{\mu\nu}(k) & = & 0\,,\nonumber\\
R_{\mu\nu}(k) \gamma^\nu & = & 0\,.
\eeqa

From the definition of $\tilde\gamma(k)$ in \Eq{gammatilde},
and assuming the on-shell condition of \Eq{onshellk},
the following relations are readily verified,
\beqa
\label{kslashgammatilde}
(\lslash{k} \pm m_\lambda)\tilde\gamma_\mu(k) & = &
(\lslash{k} \pm m_\lambda) \left(\gamma_\mu \mp \frac{k_\mu}{m_\lambda}\right)
\nonumber\\
& = & -\left(\gamma_\mu \pm \frac{k_\mu}{m_\lambda}\right)
(\lslash{k} \mp m_\lambda)
\nonumber\\
& = & -\tilde\gamma_\mu(k)(\lslash{k} \mp m_\lambda) \,.
\eeqa
These in turn imply,
\beq
(\lslash{k} \pm m_\lambda)\tilde\gamma_\mu(k)\tilde\gamma_\nu(k) =
-\tilde\gamma_\mu(k)(\lslash{k} \mp m_\lambda)\tilde\gamma_\nu(k) = 
\tilde\gamma_\mu(k)\tilde\gamma_\nu(k)(\lslash{k} \pm m_\lambda)\,,
\eeq
and can be used to show the following,
\beqa
\label{Ridentities}
(\lslash{k} \pm m_\lambda) R_{\alpha\beta}(k) & = &
R_{\alpha\beta}(k)(\lslash{k} \pm m_\lambda)\nonumber\\
& = & (\lslash{k} \pm m_\lambda)\tilde g_{\alpha\beta}(k) +
\frac{1}{3}
\left(\gamma_\alpha \pm \frac{k_\alpha}{m_\lambda}\right)
(\lslash{k} \mp m_\lambda)
\left(\gamma_\beta \pm \frac{k_\beta}{m_\lambda}\right)\,,\nonumber\\
& = &
(\lslash{k} \pm m_\lambda)\tilde g_{\alpha\beta}(k) -
\frac{1}{3}(\lslash{k} \pm m_\lambda)
\left(\gamma_\alpha \mp \frac{k_\alpha}{m_\lambda}\right)
\left(\gamma_\beta \pm \frac{k_\beta}{m_\lambda}\right)\,,\nonumber\\
& = & (\lslash{k} \pm m_\lambda)\tilde g_{\alpha\beta}(k) -
\frac{1}{3}\left(\gamma_\alpha \pm \frac{k_\alpha}{m_\lambda}\right)
\left(\gamma_\beta \mp \frac{k_\beta}{m_\lambda}\right)
(\lslash{k} \pm m_\lambda)\,,
\eeqa
which are useful in the evaluation of the various traces that appear
in the calculations.

\subsection{The spin-1/2 vertex function}

We parametrize the on-shell electromagnetic vertex function of the nucleons
in the form
\beq
\label{Gammafparam}
\Gamma^{(f)}_\mu = e\left[F_f(q^2)\gamma_\mu + G_f(q^2) P_\mu\right]\,.
\eeq
$F_f$ and $G_f$ are given by
\beqa
F_f & = & F^{(1)}_f - 2m_f F^{(2)}_f \,,\nonumber\\
G_f & = & F^{(2)}_f \,,
\eeqa
where the $F^{(i)}_f$ are the standard electromagnetic form factors,
which are defined by writing
\beq
\Gamma^{(f)}_\mu = e\left[F^{(1)}_f(q^2)\gamma_\mu +
iF^{(2)}_f(q^2)\sigma_{\mu\nu}q^\nu\right]\,,
\eeq
and which are normalized such that
\beqa
F^{(1)}_p(0) & = & 1 \,,\nonumber\\
F^{(1)}_n(0) & = & 0 \,,\nonumber\\
F^{(2)}_{f}(0) & = & \frac{\kappa_f}{2m_f} \,,
\eeqa
with $\kappa_p = 1.79$ and $\kappa_n = -1.91$.

\subsection{The spin-3/2 vertex fucntion}

As already mentioned in the Introduction, the spin-3/2 vertex function
is given by \Eq{majdiagvertex1}, with $g = 0$. For calculational purposes
we rewrite it in the form
\beq
\label{Gammaparam}
\Gamma_{\mu\alpha\beta} =
f D_{\mu\alpha\beta} + \frac{if}{m_\lambda} E_{\mu\alpha\beta} \,,
\eeq
where
\beq
\label{defD}
D_{\mu\alpha\beta} = 
\left(q^2\gamma_\mu - q_\mu\lslash{q}\right)\gamma_5\; g_{\alpha\beta}\,,
\eeq
and
\beq
\label{defE}
E_{\mu\alpha\beta} = q_\beta e_{\alpha\mu} - q_\alpha e_{\beta\mu}\,,
\eeq
with
\beq
\label{defe}
e_{\alpha\mu} = \epsilon_{\alpha\mu\nu\lambda} q^\nu Q^\lambda \,.
\eeq
We note the following obvious relations
\beq
\label{qdote}
q^\alpha e_{\alpha\mu} = Q^\alpha e_{\alpha\mu} = 0 \,,
\eeq
which in turn imply
\beq
\label{kdote}
e_{\alpha\mu} k^\alpha = e_{\alpha\mu} k^{\prime\,\alpha} = 0\,.
\eeq
From these we also obtain the following useful ones
\beq
\label{qdotge}
q_\alpha \tilde g^{\alpha\beta}(k) e_{\beta\mu} = 
q_\alpha \tilde g^{\alpha\beta}(k^\prime) e_{\beta\mu} = 0\,,
\eeq
and similarly with $\tilde g^{\alpha\beta}(k)$ replaced by
$\tilde g^{\alpha\beta}(k^\prime)$. Moreover,
by expanding the product of two epsilon symbols we obtain the formula
\beq
\label{ee}
e_{\alpha\mu} {e^\alpha}_\nu = q^2 Q_{\mu\nu} + Q^2 q_\mu q_\nu\,,
\eeq
where
\beq
\label{Qmunu}
Q_{\mu\nu} = Q_\mu Q_\nu - Q^2 g_{\mu\nu}\,,
\eeq
which appears in various places in the calculations in the appendices.

\section{The annihilation process}
\label{sec:annihilation}

\subsection{The squared amplitude}

The amplitude for the annihilation process
$\lambda(k) + \lambda(k^\prime) \rightarrow \ell(p) + \bar\ell(p^\prime)$
is given by
\beq
iM = \frac{ie}{q^2}\left[\bar v(p^\prime,s^\prime) \gamma^\mu u(p,s)\right]
\left[\bar V^\alpha(k^\prime,\sigma^\prime)
\Gamma_{\mu\alpha\beta} U^\beta(k,\sigma)\right]\,,
\eeq
where $u,v$ are the spin-1/2 Dirac spinors while $U,V$ are the spin-3/2
Rarita-Schwinger spinors, with the conventions set in
Section\ \ref{sec:preliminaries}.

The amplitude squared, averaged over the initial spins and
summed over the final spins will be expressed as the sum of three terms,
corresponding to the square of the $D$ term in \Eq{Gammaparam},
the square of the $E$ term and the interference term, respectively.
Thus,
\beq
\label{amp2assum}
\langle|M|^2\rangle = e^2 f^2 \left({\cal M} + {\cal M}^\prime +
{\cal M}^{\prime\prime}\right)\,,
\eeq
where
\beqa
\label{M2}
{\cal M} & = & \ell^{\mu\nu} L_{\mu\nu} \,,\nonumber\\
{\cal M}^\prime & = &
\left(\frac{1}{m_\lambda q^2}\right)^2
\ell^{\mu\nu} L^\prime_{\mu\nu}\,,\nonumber\\
{\cal M}^{\prime\prime} & = &
\left(\frac{1}{m_\lambda q^2}\right)
\ell^{\mu\nu} L^{\prime\prime}_{\mu\nu}\,.
\eeqa
$\ell_{\mu\nu}$ is given by
\beq
\label{ellmunu}
\ell_{\mu\nu}  = \frac{1}{4}\mbox{Tr}\,\gamma_\mu
(\lslash{p} + m_\ell)\gamma_\nu(\lslash{p}^\prime - m_\ell)\,,
\eeq
while
\beqa
\label{Lmunu}
L_{\mu\nu} & = & \frac{1}{4}\mbox{Tr}\,\gamma_\mu\gamma_5\rho^{\alpha\beta}(k)
\gamma_\nu\gamma_5 \bar\rho_{\beta\alpha}(k^\prime)\,,\\
\label{Lprimemunu}
L^\prime_{\mu\nu} & = & \frac{1}{4}E_{\mu\alpha\beta} E_{\nu\sigma\tau}
\mbox{Tr}\,\rho^{\beta\tau(k)}\bar\rho^{\sigma\alpha}(k^\prime)\,,\\
\label{Ldprimemunu}
L^{\prime\prime}_{\mu\nu} & = & -\left(\frac{i}{4}\right)
g_{\sigma\tau} E_{\nu\alpha\beta}
\mbox{Tr}\,\gamma_\mu\gamma_5
\rho^{\tau\beta}(k)\bar\rho^{\alpha\sigma}(k^\prime)\nonumber\\
&&\mbox{} +
\left(\frac{i}{4}\right)
g_{\sigma\tau} E_{\mu\alpha\beta}
\mbox{Tr}\,\gamma_\nu\gamma_5
\bar\rho^{\sigma\alpha}(k^\prime)
\rho^{\beta\tau}(k) \,,
\eeqa
where $E_{\mu\alpha\beta}$ is defined in \Eq{defE}, and
the spin-3/2 projection matrices
$\rho_{\mu\nu}(k), \bar\rho_{\mu\nu}(k^\prime)$ are given 
in \Eq{spin32poloper}.
In the above expressions for $L_{\mu\nu}$ and $L^{\prime\prime}_{\mu\nu}$
the terms of the spin-3/2 vertex function that
contain a factor of $q_\mu$ or $q_\nu$ have dropped out due to
the relation (current conservation)
\begin{equation}
\label{qdotell}
q^\mu \ell_{\mu\nu} = q^\nu \ell_{\mu\nu} = 0\,.
\end{equation}

$\ell_{\mu\nu}$ is of course trivially evaluated to yield
\beq
\label{ellmunufinal}
\ell_{\mu\nu} = p_\mu p^\prime_\nu + p^\prime_\mu p_\nu -
\frac{1}{2}q^2 g_{\mu\nu} \,.
\eeq

The evaluation of the traces that appear in the expressions for
$L_{\mu\nu},L^\prime_{\mu\nu},L^{\prime\prime}_{\mu\nu}$
is facilitated by using the identities and relations for the
RS spinors and polarizations sums given in Section\ \ref{sec:preliminaries}.
However, the calculations are involved and therefore we provide some details
in appendices\ \ref{appendixA}, \ref{appendixB} and \ref{appendixC}
respectively. The final results obtained in the appendices
can be written in the form
\beqa
\label{Lmunufinal}
L_{\mu\nu} & = & -L Q_{\mu\nu} + O(q_\mu,q_\nu)\,,\nonumber\\
L^\prime_{\mu\nu} & = & -(m_\lambda q^2)^2 L^\prime Q_{\mu\nu} +
O(q_\mu,q_\nu)\,,\nonumber\\
L^{\prime\prime}_{\mu\nu} & = & -(m_\lambda q^2) L^{\prime\prime}
Q_{\mu\nu} + O(q_\mu,q_\nu)\,,
\eeqa
where $Q_{\mu\nu}$ is defined in \Eq{Qmunu},
\beqa
\label{Lfinal}
L & = & \frac{1}{3}\left[1 +
\frac{2}{3}\left(\frac{Q^2}{2m^2_\lambda} - 1\right)^2\right]\,,\nonumber\\
L^\prime & = &
\left(\frac{5}{36}\right)\left(\frac{Q^2}{m^2_\lambda}\right)^2\,,\nonumber\\
L^{\prime\prime} & = & 
-\frac{Q^2 (Q^2 + m^2_\lambda)}{9 m^4_\lambda}\,,
\eeqa
and $O(q_\mu,q_\nu)$ stands for terms that are proportional
to $q_\mu$ and/or $q_\nu$, which do not contribute to the squared
amplitude when they are contracted with $\ell_{\mu\nu}$. Therefore,
from Eqs.\ (\ref{amp2assum}), (\ref{M2}) and (\ref{Lmunufinal}),
\beq
\label{amp2finalann}
\langle|M|^2\rangle = e^2 f^2 (L + L^\prime + L^{\prime\prime})
\left(-Q^{\mu\nu}\ell{\mu\nu}\right)\,.
\eeq

\subsection{The cross section}

We compute the cross section and the rate in the center of mass
coordinate system of the particles. The contraction formula
that appears in \Eq{amp2finalann} is readily evaluated to yield
\beq
\label{ellLcontractions}
(Q^2 g^{\mu\nu} - Q^\mu Q^\nu)\ell_{\mu\nu} = \frac{1}{2}(s - 4m^2_\lambda)sA
\eeq
where
\beq
\label{defA}
A = \left(1 + \frac{4m^2_\ell}{s}\right) +
\left(1 - \frac{4m^2_\ell}{s}\right)\cos^2\theta\,,
\eeq
with $\theta$ being the scattering angle in the CM mass frame, i.e.,
\beq
\cos\theta = \hat p\cdot\hat k\,.
\eeq
and $s = q^2$. Therefore, from \Eqs{amp2finalann}{ellLcontractions},
\beq
\label{M2final}
\langle|M|^2\rangle = e^2 f^2(L + L^\prime + L^{\prime\prime})
\frac{1}{2}(s - 4m^2_\lambda)sA\,,
\eeq
where $L, L^\prime, L^{\prime\prime}$ are given in \Eq{Lfinal},
and using \Eq{Q2ann} can be expressed in the form
\beqa
\label{Lfinalcalculations}
L & = & \frac{1}{3}\left[1 +
\frac{2}{3}\left(\frac{s}{2m^2_\lambda} - 1\right)^2\right]\,,\nonumber\\
L^\prime & = &
\frac{5}{36}\left(\frac{s}{m^2_\lambda} - 4\right)^2\,,\nonumber\\
L^{\prime\prime} & = &
-\frac{1}{9}\left(\frac{s}{m^2_\lambda} - 4\right)
\left(\frac{s}{m^2_\lambda} - 5\right)\,.
\eeqa
From the standard formula for the cross section in the CM frame,
\beq
\label{sigmadefcm}
\frac{d\sigma}{d\cos\theta} = \frac{|\vec p|}{32\pi s|\vec k|}
\langle|M|^2\rangle\,,
\eeq
we then obtain finally
\beq
\label{crosssection32ann}
\left(
\frac{d\sigma}{d\cos\theta}
\right)_{\lambda\lambda\rightarrow\ell\bar\ell} = 
\frac{r e^2 f^2}{32\pi}B(m^2_\lambda)\,,
\eeq
where we have defined
\beq
\label{defr}
r = \frac{1}{2}\left(L + L^\prime + L^{\prime\prime}\right) =
\frac{1}{6}\left(\frac{s}{2m^2_\lambda} - 1\right)^2 + \frac{1}{9}\,.
\eeq
%
%
%
%
and
\beq
\label{Bdef}
B(m^2) = 
\sqrt{(s - 4m^2)(s - 4m^2_\ell)}\,A\,,
\eeq
with $A$ given in \Eq{defA}.

\subsection{The spin-1/2 and spin-1 cases}
\label{subsec:examplesann}

For reference purposes and comparison it is useful to quote the
corresponding results for a spin-1/2 Majorana particle and for a
self-conjugate spin-1 particle, which we denote by $\chi$ and $V$ respectively.
The corresponding electromagnetic vertex functions are given
by\cite{n:nuem,np:spin1}
\beqa
\label{nuemvertex}
\Gamma^{(\chi)}_\mu & = & f (q^2\gamma_\mu - q_\mu\lslash{q})\gamma_5\,,\\
\label{spin1emvertex}
\Gamma^{(V)}_{\mu\alpha\beta} & = &
if\left(q_\beta\epsilon_{\alpha\mu\nu\lambda} -
q_\alpha\epsilon_{\beta\mu\nu\lambda}\right)q^\nu Q^\lambda\,.
\eeqa

\subsubsection{spin-1/2}

Using the $\Gamma^{(\chi)}_\mu$ vertex function,
the amplitude for $\chi\chi\rightarrow\ell\bar\ell$ is
\beq
M = ef\left[\bar u(p)\gamma^\mu v(p^\prime)\right]
\left[\bar v(k^\prime)\gamma_\mu\gamma_5 u(k)\right]\,,
\eeq
and taking the relevant traces
\beq
\label{M2nuann}
\langle|M|^2\rangle = 2e^2 f^2\left(-Q^{\mu\nu}\ell_{\mu\nu}\right) \,,
\eeq
which can be evaluated explicitly by using \Eq{ellLcontractions}.
From \Eqs{sigmadefcm}{M2nuann} the annihilation cross section
is then given by
\beq
\label{crosssectionnuann}
\left(\frac{d\sigma}{d\cos\theta}\right)_{\chi\chi\rightarrow\ell\bar\ell} =
\frac{e^2 f^2}{32\pi}B(m^2_\chi)\,,\\
\eeq
with $B$ defined in \Eq{Bdef}.

\subsubsection{spin-1}

In analogous fashion, using the $\Gamma^{(V)}_{\mu\alpha\beta}$ vertex
function the averaged squared amplitude for $VV\rightarrow \ell\bar\ell$ is
\beq
\langle|M|^2\rangle = \frac{4}{9} \frac{e^2 f^2}{s^2} V^{\mu\nu}\ell_{\mu\nu}
\eeq
where
\beq
V_{\mu\nu} = \tilde g^{\alpha\sigma}(k^\prime) \tilde g^{\beta\tau}(k)
E_{\mu\alpha\beta} E_{\nu\sigma\tau} \,,
\eeq
with $\tilde g_{\alpha\beta}(k)$ and $E_{\mu\alpha\beta}$
defined in \Eqs{gtilde}{defE}, respectively. The above expression
for $V_{\mu\nu}$ coincides with the quantity defined as $K_{\mu\nu 1}$
in \Eq{LprimeKdef}. Borrowing the result given in \Eq{LprimeKfinal}
and using \Eq{ee} we then have
\beq
V_{\mu\nu} = 
-2s^2\left(\frac{s}{4m^2_V} - 1\right) Q_{\mu\nu} + O(q_\mu,q_\nu)\,,
\eeq
and therefore
\beq
\label{M2spin1finalann}
\langle|M|^2\rangle = \frac{8}{9} e^2 f^2 \left(\frac{s}{4m^2_V} - 1\right)
\left(-Q^{\mu\nu}\ell_{\mu\nu}\right)\,,
\eeq
which can be expressed explicitly by using \Eq{ellLcontractions} once more.
Thus, from \Eqs{sigmadefcm}{M2spin1finalann} 
\beq
\label{crosssectionspin1ann}
\left(\frac{d\sigma}{d\cos\theta}\right)_{VV\rightarrow\ell\bar\ell} =
\frac{4}{9}\left(\frac{s}{4m^2_V} - 1\right)\frac{e^2 f^2}{32\pi}B(m^2_V)\,.
\eeq

\subsection{Discussion}
\label{sec:annihilationenergylimits}

The angular dependence in each of the there cases is the same,
and the total cross section in each one is proportional to
\beq
A_0 = \int\,d(\cos\theta) A = \frac{8}{3} + \frac{16 m^2_\ell}{s} \,.
\eeq
However, the energy dependence is different. For definiteness let us
assume that
\beq
m_{\lambda,\chi,V} \gg m_\ell\,,
\eeq
which is the case in most situations of practical interest.
Then $A_0 \simeq \frac{8}{3}$ and defining
\beq
B_0(m) = \int\,d(\cos\theta) B(m) \,,
\eeq
we have
\beq
B_0(m) = \frac{8}{3} \times \left\{
\begin{array}{ll}
4|\vec k|^2 & (\mbox{for} \; |\vec k| \gg m)\\[12pt]
4|\vec k|m & (\mbox{for} \; |\vec k| \ll m)
\end{array}
\right.
\eeq
where $|\vec k|$ is the magnitude of the momentum of one of the initial
particles. The relative importance of the contributions from
the two terms of the vertex function (i.e., the $D$ and $E$ terms
in \Eq{Gammaparam}) is given by
the quantities denoted $L$ and $L^\prime$ in \Eq{Lfinalcalculations}
($L^{\prime\prime}$ is the contribution from the interefence term).
Since their behaviour is
quite different in the high or the low energy limit let us consider
them separately. 

\subsubsection{High energy limit}

In the limit $s \simeq 4|\vec k|^2 \gg 4m^2_\lambda$ the $D$ and $E$ terms
 give a comparable contribution,
\beqa
L & \simeq & \frac{s^2}{18m^4_\lambda}\,,\nonumber\\
L^\prime & \simeq & \frac{5s^2}{36 m^4_\lambda}\,,\nonumber\\
L^{\prime\prime} & \simeq & -\frac{s^2}{9m^4_\lambda}\,.
\eeqa
The total cross section in this limit is then
\beq
\sigma_{\lambda\lambda\rightarrow\ell\bar\ell} \simeq 
\left(\frac{s^2}{24m^4_\lambda}\right)\frac{e^2 f^2 s}{32\pi} \,,
\eeq
whereas
\beqa
\sigma_{\chi\chi\rightarrow\ell\bar\ell} & \simeq & \frac{e^2 f^2 s}{32\pi}\,,
\nonumber\\
\sigma_{VV\rightarrow\ell\bar\ell} & \simeq  &
\left(\frac{s}{9m_V}\right)\frac{e^2 f^2 s}{32\pi} \,.
\eeqa

\subsubsection{Low energy limit}

In the limit $|\vec k|^2 \ll m^2_\lambda$\,,
\beqa
L & \simeq & \frac{5}{9}\,,\nonumber\\
L^\prime & = & \frac{20|\vec k|^4}{9m^4_\lambda}\,,\nonumber\\
L^{\prime\prime} & = & \frac{4|\vec k|^2}{9m^2_\lambda}\,,
\eeqa
so that the $E$ term gives a negligible contribution compared with the
$D$ term. The total cross section is then
\beq
\sigma_{\lambda\lambda\rightarrow\ell\bar\ell} \simeq 
\left(\frac{5}{18}\right)\frac{e^2 f^2 |\vec k| m_\lambda}{3\pi} \,,
\eeq
which is very similar to the result for the spin-1/2 case,
\beq
\sigma_{\chi\chi\rightarrow\ell\bar\ell} \simeq 
\frac{e^2 f^2 |\vec k| m_\chi}{3\pi} \,.
\eeq
Thus, for example, in the context of the calculation of the cosmological
relic abundance of the $\lambda$ particles, following the steps outlined in
\Ref{scherrer1} in the spin-1/2 case to calculate the thermal average of
the total rate $v_{rel}\sigma$, we obtain in this case
\beq
\langle v_{rel}\sigma_{\lambda\lambda\rightarrow\ell\bar\ell}\rangle =
\frac{5 e^2 f^2 m_\lambda T}{18 \pi}\,,
\eeq
which has the same temperature dependence as the in spin-1/2 case
\beq
\langle v_{rel}\sigma_{\chi\chi\rightarrow\ell\bar\ell}\rangle =
\frac{e^2 f^2 m_\chi T}{\pi}\,.
\eeq

\section{The scattering process}
\label{sec:scattering}

Here we consider the scattering process
$\lambda(k) + f(p) \rightarrow \lambda(k^\prime) + f(p^\prime)$.
In analogy with \Eq{amp2assum}, the squared amplitude in this case can
be expressed in the form
\beq
\label{amp2assumscatt}
\langle|M|^2\rangle = 2e^2 f^2 \left({\cal M} + {\cal M}^\prime +
{\cal M}^{\prime\prime}\right)\,,
\eeq
where
\beqa
\label{M2scatt}
{\cal M} & = & f^{\mu\nu} \tilde L_{\mu\nu} \,,\nonumber\\
{\cal M}^\prime & = &
\left(\frac{1}{m_\lambda q^2}\right)^2
f^{\mu\nu} \tilde L^\prime_{\mu\nu}\,,\nonumber\\
{\cal M}^{\prime\prime} & = &
\left(\frac{1}{m_\lambda q^2}\right)
f^{\mu\nu} \tilde L^{\prime\prime}_{\mu\nu}\,.
\eeqa
With the convention given in \Eq{Gammafparam} for the
electromagnetic vertex function of the spin-1/2 fermion,
$f_{\mu\nu}$ is given by
\beq
\label{fmunu}
f_{\mu\nu}  = \frac{1}{4}\mbox{Tr}\,(F_f\gamma_\mu + G_f P_\mu)
(\lslash{p} + m_f)(F_f\gamma_\nu + G_f P_\nu)(\lslash{p}^\prime + m_f)\,,
\eeq
which is straightforwardly evaluated to yield
\beqa
f_{\mu\nu} & = & F^2_f \left[p_\mu p^\prime_\nu + p^\prime_\mu p_\nu +
\frac{1}{2}q^2 g_{\mu\nu}\right] +
\left[G^2_f\left(2 m^2_f - \frac{1}{2}q^2\right) + 2m_f F_f G_f\right]
P_\mu P_\nu \,.
\eeqa
On the other hand, $\tilde L_{\mu\nu}$, $\tilde L^\prime_{\mu\nu}$
and $\tilde L^{\prime\prime}_{\mu\nu}$
are given by expressions analogous to those for
$L$, $L^\prime$ and $L^{\prime\prime}$ in \Eq{Lmunu}, respectively,
but with the replacement
$\bar\rho_{\mu\nu}(k^\prime)\rightarrow\rho_{\mu\nu}(k^\prime)$,
which can be expressed conveniently by the substitution rule
\beqa
\tilde L_{\mu\nu} & = & -L_{\mu\nu}(k^\prime\rightarrow -k^\prime)
\nonumber\\
\tilde L^\prime_{\mu\nu} & = &
-L^\prime_{\mu\nu}(k^\prime\rightarrow -k^\prime)
\nonumber\\
\tilde L^{\prime\prime}_{\mu\nu} & = & 
-L^{\prime\prime}_{\mu\nu}(k^\prime\rightarrow -k^\prime)\,.
\eeqa
Thus, from \Eq{Lmunufinal}, we have
\beqa
\tilde L_{\mu\nu} & = & \tilde L\, Q_{\mu\nu} +
O(q_\mu,q_\nu)\,,\nonumber\\
\tilde L^\prime_{\mu\nu} & = & (m_\lambda q^2)^2 \tilde L^\prime\, Q_{\mu\nu} +
O(q_\mu,q_\nu)\,,\nonumber\\
\tilde L^{\prime\prime}_{\mu\nu} & = &
(m_\lambda q^2) \tilde L^{\prime\prime}\, Q_{\mu\nu} + O(q_\mu,q_\nu)\,,
\eeqa
where
\beqa
\label{tildeL}
\tilde L & = & \frac{1}{3}\left[1 +
\frac{2}{3}\left(1- \frac{t}{2m^2_\lambda}\right)^2\right]\,,\nonumber\\
\tilde L^\prime & = &
\frac{5}{36}\left(4 - \frac{t}{m^2_\lambda}\right)^2\,,\nonumber\\
\tilde L^{\prime\prime} & = &
-\frac{1}{9}\left(4 - \frac{t}{m^2_\lambda}\right)
\left(5 - \frac{t}{m^2_\lambda}\right)\,.
\eeqa
In writing \Eq{tildeL} we have used the kinematic relations
given in \Eq{Q2scatt}. The expression in \Eq{amp2assumscatt} for the squared
amplitude then yields
\beq
\langle|M|^2\rangle = 2e^2 f^2 (\tilde L + \tilde L^\prime +
\tilde L^{\prime\prime}) Q^{\mu\nu}f_{\mu\nu} \,.
\eeq
The final ingredient to compute the cross section is the contraction
$Q^{\mu\nu} f_{\mu\nu}$. Using $Q^{\mu\nu} (p - p^\prime)_\nu = 0$
we have, on one hand
\beq
Q^{\mu\nu} P_\mu P_\nu = 4 Q^{\mu\nu} p_\mu p_\nu\,,
\eeq
and on the other hand 
\beq
Q^{\mu\nu}\left[p_\mu p^\prime_\nu + p^\prime_\mu p_\nu +
\frac{1}{2}q^2 g_{\mu\nu}\right] = 2 Q^{\mu\nu} p_\mu p_\nu +
t(t - 4m^2_\lambda) \,,
\eeq
where $Q^{\mu\nu} p_\mu p_\nu$ is 
in turn is straightforwardly evaluated and it can be expressed  in the form
\beq
Q^{\mu\nu} p_\mu p_\nu = 
(s - m^2_f - m^2_\lambda)^2 + st - 4m^2_f m^2_\lambda\,,
\eeq
where we have introduced
\beq
s = (p + k)^2 \,.
\eeq
Putting these together, we then obtain
\beq
Q^{\mu\nu} f_{\mu\nu} = C(m^2_\lambda) \,,
\eeq
where
\beq
\label{C}
C(m^2) =  t(t - 4m^2)F^2_f +
\left[2(s - m^2_f - m^2)^2 + 2st - 8m^2_f m^2\right]
\left[F^2_f + G^2_f(4m^2_f - t) + 4m_f F_f G_f\right] \,.
\eeq
From the standard formula
\beq
\frac{d\sigma}{dt} = \frac{1}{64\pi m^2_f |\vec k|^2}\langle|M|^2\rangle\,,
\eeq
the final result for the cross section is
\beq
\label{crosssection32scatt}
\left(\frac{d\sigma}{dt}\right)_{\lambda f\rightarrow \lambda f} =
\frac{\tilde r e^2 f^2}{32\pi m^2_f |\vec k|^2}C(m^2_\lambda)\,.
\eeq
where
\beq
\label{deftilder}
\tilde r = \tilde L + \tilde L^\prime + \tilde L^{\prime\prime} =
\frac{1}{3}\left(1 - \frac{t}{2m^2_\lambda}\right)^2 + \frac{2}{9}\,.
\eeq

Again it is useful to quote the corresponding results
for the spin-1/2 Majorana ($\chi$) and the self-conjugate spin-1
particle ($V$), for which the electromagnetic vertex functions are given
by \Eqs{nuemvertex}{spin1emvertex} respectively. The results can be obtained
by using as guidance the discussion in Section\ \ref{subsec:examplesann}.
In particular the results given in \Eq{M2nuann} for $\chi$
and in \Eq{M2spin1finalann} for $V$ can be reused,
remembering that in the present case $\ell_{\mu\nu}\rightarrow f_{\mu\nu}$
and by properly taking into account the changes due to the $t$-channel
kinematics and the difference in the numerical factors that arise
from the averaging (sum) over the initial (final) polarizations.
Thus,
\beq
\label{crosssectionnuscatt}
\left(\frac{d\sigma}{dt}\right)_{\chi f\rightarrow \chi f} = 
\frac{e^2 f^2}{32\pi m^2_f |\vec k|^2}C(m^2_\chi)\,,
\eeq
is the cross section for
$\chi + f \rightarrow \chi + f$, with the $\chi$ electromagnetic
vertex function given by \Eq{nuemvertex}.
Similarly for $V$, with the vertex function given by \Eq{spin1emvertex},
the cross section for $V + f \rightarrow V + f$ is
\beq
\label{crosssectionspin1scatt}
\left(\frac{d\sigma}{dt}\right)_{Vf\rightarrow Vf} =
\frac{2}{3}\left(1 - \frac{t}{4m^2_V}\right)
\frac{e^2 f^2}{32\pi m^2_f |\vec k|^2}C(m^2_V)\,.
\eeq

\subsection{Discussion}

\Eqs{C}{crosssection32scatt} can be used to determine the scattering
cross section off a high $Z$ nucleus $N$, which is a relevant quantity
in the context of direct dark matter detection experiments. In this
context, using the subscript $N$ in place of $f$ wherever the latter
appears, we set
\beqa
G_N \rightarrow & 0\,,\nonumber\\
F_N \rightarrow & Z \,.
\eeqa
Denoting the kinetic energy of the recoil nucleus by $\epsilon^\prime$, 
using the notation given in \Eq{componentnotation} for the components
of the momentum vectors,
\beq
E^\prime = m_N + \epsilon^\prime \,,
\eeq
and $t$ and $s$ are given by
\beqa
t & = & -2m_N \epsilon^\prime\,,\nonumber\\
s & = & m^2 + m^2_N + 2m_N\omega \,,
\eeqa
in terms of the variables in the rest frame of the initial nucleus.
$C(m^2)$ can then be written in the form
\beq
C(m^2) = 4m^2_N Z \left[C_1(m^2) + C_2(m^2)\right] \,,
\eeq
where
\beqa
C_1(m^2) & = & 2(\omega^2 - m^2) + \frac{\epsilon^\prime}{m_N}
\left(m^2 - m^2_N - 2mm_N\right)\,,\nonumber\\
C_2(m^2) & = & \epsilon^\prime\left[\epsilon^\prime - (\omega - m)\right]\,.
\eeqa
On the other hand, the factor  $\tilde r$ defined in \Eq{deftilder}
is given by
\beq
\label{tilder2}
\tilde r = \frac{2}{9} + \frac{1}{3}
\left(1 + \frac{m_N\epsilon^\prime}{m^2_\lambda}\right)^2 \,.
\eeq
In this process the relative contributions from the $D$ and $E$ terms
of the vertex function, represented by the terms $\tilde L, \tilde L^\prime$
in \Eq{deftilder}, are comparable. The differential cross section
is then given as a function of $\epsilon^\prime$ by
\beq
\left(\frac{d\sigma}{d\epsilon^\prime}\right)_{\lambda N\rightarrow\lambda N}
= \frac{e^2 f^2}{16\pi m_N |\vec k|^2} \tilde r C(m^2_\lambda) \,.
\eeq
We now consider specifically the non-relativistic limit since this is a
particularly relevant situation.

\subsubsection*{Non-relativistic limit}

Denoting by $v$ the initial velocity of the $\lambda$ particle,
in the non-relativistic limit $C_2 = O(v^4)$, and therefore
\beq
C(m^2) = 2 v^2 +
\frac{\epsilon^\prime}{m_N}\left(m^2 - m^2_N - 2mm_N\right) + O(v^4)\,.
\eeq
The differential cross section is then given by
\beq
\label{scattsigmanr}
\left(\frac{d\sigma}{d\epsilon^\prime}\right)_{\lambda N\rightarrow\lambda N}
= \frac{e^2 f^2}{16\pi m_N m^2_\lambda} \tilde r\left[
2 + \frac{\epsilon^\prime}{m_N v^2}
\left(m^2_\lambda - m^2_N - 2m_\lambda m_N\right)\right] \,,
\eeq
where $0 \le \epsilon^\prime \le \epsilon^\prime_{max}$ with
\beq
\epsilon^\prime_{max} = \frac{2m_N m^2_\lambda v^2}{(m_N + m_\lambda)^2} \,.
\eeq
Using this in \Eq{tilder2},
\beq
\tilde r_{max} = \frac{2}{9} + \frac{1}{3}
\left(1 + \frac{2 m^2_N v^2}{(m_N + m^2_\lambda)^2}\right)^2 \,,
\eeq
and therefore we can put $\tilde r \simeq \frac{5}{9} + O(v^2)$
in \Eq{scattsigmanr} within the approximations made there.
The coresponding cross section for the spin-1/2 $\chi$-particle is given by
\Eq{scattsigmanr}, with the substitution $\tilde r \rightarrow 1$
and of course $m_\lambda\rightarrow m_\chi$.

\section{Conclusions}
\label{sec:conclusions}

In this work we have calculated the cross section for the electromagnetic
pair annihilation of a spin-3/2 Majorana particle ($\lambda$)
into charged leptons, and for the electromagnetic scattering of
such particle off a nucleon. The calculations are based on the general
structure of electromagnetic vertex function obtained in \Ref{n:spin32m}
for the spin-3/2 Majorana particle.

The main results of our calculations are the expressions for the
annihilation and scattering cross sections, given in
\Eqs{crosssection32ann}{crosssection32scatt}, respectively. Comparing
with the corresponding formulas for the spin-1/2 case
given in \Eqs{crosssectionnuann}{crosssectionnuscatt}, reveal
certain similarities. For example the angular dependence of
annihilation cross section is the same in the spin-1/2 and spin-3/2 cases,
and at low energies, $s \simeq 4m^2_\lambda$, both of them are proportional
to the momentum $|\vec k|$, which is a frequently-quoted result
in the context of the spin-1/2 case that is known to be due to the
axial vector nature of the $\gamma_\mu\gamma_5$ coupling\cite{k:nucpt}.
In Section\ \ref{sec:annihilationenergylimits} we considered in more detail
this limit and gave the formulas that can be useful for example
in the context of the applications similar to those considered in
Refs.\ \cite{scherrer1,scherrer2} for the spin-1/2 case.

However there are important differences as well. The differences between
the cross sections appear in the overall factors denoted by $r$ and
$\tilde r$ in \Eqs{crosssection32ann}{crosssection32scatt}
for the annihilation and scattering cross sections
respectively, which are given explicitly in \Eqs{defr}{deftilder}.
Thus, the energy dependence of the cross sections for
the spin-3/2 and spin-1/2 cases are generally different. In addition,
due to the dependence of $\tilde r$ on $t$,
the angular dependence of the scattering cross section
is also different in the spin-3/2 and spin-1/2 cases.

Two other processes which could be of interest are the annihilation
into photons and scattering off photons,
\beqa
\lambda + \lambda & \rightarrow & \gamma + \gamma\,,\nonumber\\
\lambda + \gamma & \rightarrow & \lambda + \gamma\,.
\eeqa
In connection with this we wish to make the following remark.
In the spin-1/2 case, the amplitudes for the analogous processes
\beqa
\chi + \chi & \rightarrow & \gamma + \gamma\,,\nonumber\\
\chi + \gamma & \rightarrow & \chi + \gamma\,,
\eeqa
vanish at the tree-level. This follows from the fact
that in both cases the amplitude contains a factor of the form
\beq
\epsilon^\mu (q^2\gamma_\mu - \lslash{q} q_\mu)\gamma_5 \,,
\eeq
one such factor for each of the photons in the process in question,
and such factors vanish for an on-shell photon. In contrast,
in the spin-3/2 vertex function [\Eq{Gammafparam}]
the $D$ term [\Eq{defD}] is of similar form and
also vanishes for on-shell photons,
but the $E$ term [\Eq{defE}] does not vanish and consequently the tree-level
amplitude is non-zero. This crucial distinction
between the spin-1/2 and spin-3/2 cases can have interesting
implications in some of the physical application contexts that
we have mentioned in the Introduction. We plan to carry out
the corresponding calculations for those radiative process and
present them in the near future.

\begin{acknowledgments}
The  work of S. S. is partially supported by DGAPA-UNAM (Mexico) Project
No. IN103812.
\end{acknowledgments}
\appendix

\section{The $L_{\mu\nu}$ trace}
\label{appendixA}

Using \Eq{spin32poloper}, the expression for $L_{\mu\nu}$ given in \Eq{Lmunu}
can be rewritten in the form
\beq
\label{EvalL1}
L_{\mu\nu} = \frac{1}{4}\mbox{Tr}\,\gamma_\mu (\lslash{k} + m_\lambda)
R^{\alpha\beta}(k)\gamma_\nu R_{\beta\alpha}(k^\prime)
(\lslash{k}^\prime + m_\lambda)\,.
\eeq
With the help of the relations given in \Eq{Ridentities},
\Eq{EvalL1} can be expressed in the form
\begin{equation}
L_{\mu\nu} = \sum^4_{i = 1} \frac{1}{4}\mbox{Tr}\;L^{(i)}_{\mu\nu}
\end{equation}
where
\begin{eqnarray}
L^{(1)}_{\mu\nu} & = &
\tilde g^{\alpha\beta}(k)\tilde g_{\beta\alpha}(k^\prime)
\gamma_\mu(\lslash{k} + m_\lambda)\gamma_\nu(\lslash{k}^\prime + m_\lambda)
\,,\nonumber\\
L^{(2)}_{\mu\nu} & = & -\frac{1}{3}\gamma_\mu(\lslash{k} + m_\lambda)
\left(\gamma^\alpha - \frac{k^\alpha}{m_\lambda}\right)
\left(\gamma^\beta + \frac{k_\beta}{m_\lambda}\right)
\tilde g_{\beta\alpha}(k^\prime)\gamma_\nu(\lslash{k}^\prime + m_\lambda)
\,,\nonumber\\
L^{(3)}_{\mu\nu} & = & -\frac{1}{3}\gamma_\mu(\lslash{k} + m_\lambda)
\gamma_\nu \tilde g^{\alpha\beta}(k)
\left(\gamma_\beta + \frac{k^\prime_\beta}{m_\lambda}\right)
\left(\gamma_\alpha - \frac{k^\prime_\alpha}{m_\lambda}\right)
(\lslash{k}^\prime + m_\lambda)
\,,\nonumber\\
L^{(4)}_{\mu\nu} & = & \frac{1}{9}
\left(\gamma^\alpha + \frac{k^{\prime\alpha}}{m_\lambda}\right)
\gamma_\mu
\left(\gamma_\alpha + \frac{k_\alpha}{m_\lambda}\right)
(\lslash{k} - m_\lambda)
\left(\gamma^\beta + \frac{k^\beta}{m_\lambda}\right)
\gamma_\nu
\left(\gamma_\beta + \frac{k^\prime_\beta}{m_\lambda}\right)
(\lslash{k}^\prime - m_\lambda)\,.
\end{eqnarray}
In writing $L^{(4)}_{\mu\nu}$ we have used the relations in
\Eq{kslashgammatilde} together with the cyclic property of the trace.

$L^{(1)}_{\mu\nu}$ is easily shown to be given by
\beq
\label{L1final}
L^{(1)}_{\mu\nu} = \left[2 + \left(\frac{k\cdot k}{m^2_\lambda}\right)^2\right]
\gamma_\mu(\lslash{k} + m_\lambda)\gamma_\nu(\lslash{k}^\prime + m_\lambda)
\eeq
while using the relations
\begin{equation}
(\lslash{k} + m_\lambda)\lslash{k} =
\lslash{k}(\lslash{k} + m_\lambda) = m_\lambda(\lslash{k} + m_\lambda)
\end{equation}
and the analogous relations with $k \rightarrow k^\prime$,
$L^{(2,3)}_{\mu\nu}$ are straightforwardly reduced to
\begin{equation}
\label{L23final}
L^{(3)}_{\mu\nu} = L^{(2)}_{\mu\nu} = -\frac{1}{3} L^{(1)}_{\mu\nu}\,.
\end{equation}
For $L^{(4)}_{\mu\nu}$, we first note the relation
\beq
(\lslash{k}^\prime - m_\lambda)
\left(\gamma^\alpha + \frac{k^{\prime\alpha}}{m_\lambda}\right)
\gamma_\mu
\left(\gamma_\alpha + \frac{k_\alpha}{m_\lambda}\right)
(\lslash{k} - m_\lambda) =
(\lslash{k}^\prime - m_\lambda)\left[
\frac{k\cdot k^\prime}{m^2_\lambda}\gamma_\mu +
\frac{2(k + k^\prime)_\mu}{m_\lambda}\right](\lslash{k} - m_\lambda)
\eeq
and the corresponding relation with $k \leftrightarrow k^\prime$.
These relations, together with the cyclic property of the trace
then imply
\beqa
\label{L4final}
\mbox{Tr}\; L^{(4)}_{\mu\nu} & = & \frac{1}{9}\mbox{Tr}\;
\left[\frac{k\cdot k^\prime}{m^2_\lambda}\gamma_\mu +
\frac{2(k + k^\prime)_\mu}{m_\lambda}\right](\lslash{k} - m_\lambda)
\left[\frac{k\cdot k^\prime}{m^2_\lambda}\gamma_\nu +
\frac{2(k + k^\prime)_\nu}{m_\lambda}\right](\lslash{k}^\prime - m_\lambda)
\nonumber\\
& = & \frac{1}{9}\left(\frac{k\cdot k^\prime}{m^2_\lambda}\right)^2
\mbox{Tr}\; \gamma_\mu (\lslash{k} - m_\lambda)
\gamma_\nu(\lslash{k}^\prime - m_\lambda) + \tilde L^{(4)}_{\mu\nu} \,,
\end{eqnarray}
where $\tilde L^{(4)}_{\mu\nu}$ contains one factor of $(k + k^\prime)_\mu$
and/or $(k + k^\prime)_\nu$.
Evaluating the remaining trace in \Eqs{L1final}{L4final},
and using \Eq{L23final}, we finally obtain
\begin{equation}
\label{Lmunucalculated}
L_{\mu\nu} = \frac{2}{3}\left[1 +
\frac{2}{3}\left(\frac{k\cdot k^\prime}{m^2_\lambda}\right)^2\right]
\left[k_\mu k^\prime_\nu + k^\prime_\mu k_\nu -
\left(k\cdot k^\prime - m^2_\lambda\right)g_{\mu\nu}\right] +
\tilde L^{(4)}_{\mu\nu} \,,
\end{equation}
can be written in the form given in \Eq{Lmunufinal}
by using \Eqs{qdefann}{Qdefann}
to express $k$ and $k^\prime$ in terms of $q$ and $Q$,
and using the relation
\beq
\label{kkprimetoQmunu}
k_\mu k^\prime_\nu + k^\prime_\mu k_\nu -
\left(k\cdot k^\prime - m^2_\lambda\right)g_{\mu\nu} =
-\frac{1}{2}Q_\mu Q_\nu + \frac{1}{2}Q^2 g_{\mu\nu} + \frac{1}{2}q_\mu q_\nu
\eeq

\section{The $L^\prime_{\mu\nu}$ trace}
\label{appendixB}

Using \Eq{spin32poloper} and the relations given in \Eq{kslashgammatilde},
we rewrite the expression for \Eq{Lprimemunu} in the form
\beq
L^\prime_{\mu\nu} = \frac{1}{4}
\mbox{Tr}\,(\lslash{k} + m_\lambda) K_{\mu\nu}
(\lslash{k}^\prime - m_\lambda)
\eeq
where
\beqa
K_{\mu\nu} & = & E_{\mu\alpha\beta} E_{\nu\sigma\tau}
\left[\tilde g^{\beta\tau}(k) -
\frac{1}{3}\left(\gamma^\beta - \frac{k^\beta}{m_\lambda}\right)
\left(\gamma^\tau + \frac{k^\tau}{m_\lambda}\right)\right]
\nonumber\\
&&\mbox{}\times
\left[\tilde g^{\sigma\alpha}(k^\prime) -
\frac{1}{3}\left(\gamma^\sigma - \frac{k^{\prime\,\sigma}}{m_\lambda}\right)
\left(\gamma^\alpha + \frac{k^{\prime\,\alpha}}{m_\lambda}\right)\right]\,.
\eeqa
We express $K_{\mu\nu}$ in the form
\beq
K_{\mu\nu} = \sum^4_{i = 1}K_{\mu\nu\, i}\,,
\eeq
where
\beqa
\label{LprimeKdef}
K_{\mu\nu 1} & = & E_{\mu\alpha\beta} E_{\nu\sigma\tau}\,
\tilde g^{\beta\tau}(k)\, \tilde g^{\sigma\alpha}(k^\prime)\,,\nonumber\\
K_{\mu\nu 2} & = & E_{\mu\alpha\beta} E_{\nu\sigma\tau}
\tilde g^{\beta\tau}(k)
\left(\frac{-1}{3}\right)
\left(\gamma^\sigma - \frac{k^{\prime\,\sigma}}{m_\lambda}\right)
\left(\gamma^\alpha + \frac{k^{\prime\,\alpha}}{m_\lambda}\right)\,,
\nonumber\\
K_{\mu\nu 3} & = & E_{\mu\alpha\beta} E_{\nu\sigma\tau}
\left(\frac{-1}{3}\right)\tilde g^{\sigma\alpha}(k^\prime)
\left(\gamma^\beta - \frac{k^{\prime\,\beta}}{m_\lambda}\right)
\left(\gamma^\tau + \frac{k^{\prime\,\tau}}{m_\lambda}\right)\,,\nonumber\\
K_{\mu\nu 4} & = & \frac{1}{9} E_{\mu\alpha\beta} E_{\nu\sigma\tau}
\left(\gamma^\beta - \frac{k^\beta}{m_\lambda}\right)
\left(\gamma^\tau + \frac{k^\tau}{m_\lambda}\right)
\left(\gamma^\sigma - \frac{k^{\prime\,\sigma}}{m_\lambda}\right)
\left(\gamma^\alpha + \frac{k^{\prime\,\alpha}}{m_\lambda}\right)\,.
\eeqa
and consider each term separately.

$K_{\mu\nu\,1}$ is expressed in the form
\beq
K_{\mu\nu\,1} = K_{\mu\nu 1a} + K_{\mu\nu 1b} +
K_{\mu\nu 1c} + K_{\mu\nu 1d} \,,
\eeq
where
\beqa
K_{\mu\nu 1a} & = & E_{\mu\alpha\beta} E_{\nu\sigma\tau}\,
g^{\beta\tau}\, g^{\sigma\alpha}\,,\nonumber\\
K_{\mu\nu 1b} & = & E_{\mu\alpha\beta} E_{\nu\sigma\tau} g^{\beta\tau}
\left(-\frac{k^{\prime\,\sigma} k^{\prime\,\alpha}}{m^2_\lambda}\right)
\nonumber\\
K_{\mu\nu 1c} & = & E_{\mu\alpha\beta} E_{\nu\sigma\tau} g^{\sigma\alpha}
\left(-\frac{k^\beta k^\tau}{m^2_\lambda}\right)
\nonumber\\
K_{\mu\nu 1d} & = & E_{\mu\alpha\beta} E_{\nu\sigma\tau}
\left(-\frac{k^\beta k^\tau}{m^2_\lambda}\right)
\left(-\frac{k^{\prime\,\sigma} k^{\prime\,\alpha}}{m^2_\lambda}\right)
\eeqa
For $K_{\mu\nu 1a}$ we obtain
\beq
K_{\mu\nu 1a} = 2 q^2 e_{\alpha\mu} {e^\alpha}_\nu\,,
\eeq
where $e_{\alpha\mu} {e^\alpha}_\nu$ is given in \Eq{ee}. Similarly,
\beq
K_{\mu\nu 1c} = K_{\mu\nu 1b} = -\left(\frac{q^2}{2m_\lambda}\right)^2
e_{\alpha\mu} {e^\alpha}_\nu\,
\eeq
where we have used the relations
\beq
\label{kErel}
E_{\mu\alpha\beta} k^\beta = E_{\mu\alpha\beta}k^{\prime\,\beta} = 
\frac{1}{2}q^2 e_{\alpha\mu} \,.
\eeq
Finally,
\beq
K_{\mu\nu 1d} = 0\,,
\eeq
which follows by using \Eq{kErel}, together with \Eq{kdote}.
Summarizing,
\beq
\label{LprimeKfinal}
K_{\mu\nu 1} = \left(\frac{Q^2 q^2}{2 m^2_\lambda}\right)
e_{\alpha\mu} {e^\alpha}_\nu \,.
\eeq
Using the trace formula
\beq
\label{tracescalar}
\mbox{Tr}\,(\lslash{k} + m_\lambda)(\lslash{k}^\prime - m_\lambda) =
- 2Q^2\,,
\eeq
we then obtain
\beq
\label{K1final}
\frac{1}{4}\mbox{Tr}\,(\lslash{k} + m_\lambda)K_{\mu\nu 1}
(\lslash{k}^\prime - m_\lambda) =
-\frac{Q^2}{2}\left(\frac{Q^2 q^2}{2 m^2_\lambda}\right)
e_{\alpha\mu} {e^\alpha}_\mu \,,
\eeq
where $e_{\alpha\mu} {e^\alpha}_\mu$ is given in \Eq{ee}.

%
%

Expanding it out the expression for $K_{\mu\nu 2}$ given in \Eq{LprimeKdef},
\beqa
K_{\mu\nu 2} & = & \tilde g^{\beta\tau}(k) \left(\frac{-1}{3}\right)\left[
(E_{\nu\sigma\tau}\gamma^\sigma)(E_{\mu\alpha\beta}\gamma^\alpha) -
(E_{\mu\alpha\beta}\gamma^\alpha)
(E_{\nu\sigma\tau}\frac{k^{\prime\,\sigma}}{m_\lambda})
\right.\nonumber\\
&&\mbox{} +
\left.
(E_{\mu\alpha\beta}\frac{k^{\prime\,\alpha}}{m_\lambda})
(E_{\nu\sigma\tau}\gamma^\sigma) -
(E_{\mu\alpha\beta}\frac{k^{\prime\,\alpha}}{m_\lambda})
(E_{\nu\sigma\tau}\frac{k^{\prime\,\sigma}}{m_\lambda})
\right]
\eeqa
In the second and the third terms we will use the relation
\beq
\label{Egammasandwiched}
(\lslash{k} + m_\lambda)E_{\mu\alpha\beta}\gamma^\beta
(\lslash{k}^\prime - m_\lambda) =
(\lslash{k} + m_\lambda)
\left[-iq_\alpha(q^2\gamma_\mu - q_\mu\lslash{q})\gamma_5\right]
(\lslash{k}^\prime - m_\lambda)\,,
\eeq
which is proven as follows. From the definition of $E_{\mu\alpha\beta}$,
\beq
E_{\mu\alpha\beta}\gamma^\beta = \lslash{q} e_{\alpha\mu} -
q_\alpha e_{\beta\mu}\gamma^\beta \,,
\eeq
and therefore
\beq
\label{defEgammasandwiched}
(\lslash{k} + m_\lambda)E_{\mu\alpha\beta}\gamma^\beta
(\lslash{k}^\prime - m_\lambda) = -(\lslash{k} + m_\lambda)
q_\alpha e_{\beta\mu}\gamma^\beta(\lslash{k}^\prime - m_\lambda)\,.
\eeq
By means of the identity
\beq
\gamma_\mu\gamma_\lambda\gamma_\rho = (g_{\mu\lambda} g_{\rho\beta}
- g_{\mu\rho} g_{\lambda\beta} + g_{\mu\beta} g_{\lambda\rho})\gamma^\beta
+ i\epsilon_{\mu\lambda\rho\beta} \gamma^\beta\gamma_5 \,,
\eeq
together with the relation
\beq
(\lslash{k} + m_\lambda)
\gamma_\mu\lslash{q}\lslash{Q}\gamma_5
(\lslash{k}^\prime - m_\lambda) = 
(\lslash{k} + m_\lambda)
(q^2\gamma_\mu - q_\mu\lslash{q}\gamma_5 - Q_\mu\lslash{q})\gamma_5
(\lslash{k}^\prime - m_\lambda)\,,
\eeq
we obtain
\beq
\label{egammasandwiched1}
(\lslash{k} + m_\lambda)e_{\beta\mu}\gamma^\beta
(\lslash{k}^\prime - m_\lambda) = (\lslash{k} + m_\lambda)
i(q^2\gamma_\mu - q_\mu\lslash{q})\gamma_5
(\lslash{k}^\prime - m_\lambda)\,,
\eeq
and using this in \Eq{defEgammasandwiched} proves the identity quoted in
\Eq{Egammasandwiched}. We note here that the analogous relations
\beqa
\label{egammasandwiched2}
(\lslash{k} - m_\lambda)e_{\beta\mu}\gamma^\beta
(\lslash{k}^\prime + m_\lambda) & = & (\lslash{k} - m_\lambda)
i(q^2\gamma_\mu - q_\mu\lslash{q})\gamma_5
(\lslash{k}^\prime + m_\lambda)\nonumber\\
(\lslash{k}^\prime \pm m_\lambda)e_{\beta\mu}\gamma^\beta
(\lslash{k} \mp m_\lambda) & = & (-1)(\lslash{k}^\prime \pm m_\lambda)
i(q^2\gamma_\mu - q_\mu\lslash{q})\gamma_5
(\lslash{k} \mp m_\lambda)\,,\nonumber\\
\eeqa
follow in similar fashion.

With the help of \Eqs{Egammasandwiched}{kErel} it follows that,
when $K_{\mu\nu 2}$ is sandwiched between $(\lslash{k} + m_\lambda)$
and $(\lslash{k}^\prime - m_\lambda)$, the second and third term vanish
while the first and fourth terms give
\beqa
\label{K2sandwiched}
(\lslash{k} + m_\lambda)K_{\mu\nu 2}(\lslash{k}^\prime - m_\lambda) & = &
(\lslash{k} + m_\lambda)\left(\frac{-1}{3}\right)
\left[
\tilde g^{\beta\tau}(k)(E_{\nu\sigma\tau}\gamma^\sigma)
(E_{\mu\alpha\beta}\gamma^\alpha) - 
\left(\frac{q^2}{2m_\lambda}\right)^2 e_{\mu\alpha} {e_\nu}^\alpha
\right](\lslash{k}^\prime - m_\lambda)\,.
\eeqa
The term $e_{\mu\alpha} {e_\nu}^\alpha$ is reduced by means of \Eq{ee},
while using \Eq{defE} the first term can be expressed in the form
\beqa
\tilde g^{\beta\tau}(k)(E_{\nu\sigma\tau}\gamma^\sigma)
(E_{\mu\alpha\beta}\gamma^\alpha) & = &
q^2 e_{\tau\nu} e_{\beta\mu} g^{\beta\tau} +
\left(q_\beta q_\tau\tilde g^{\beta\tau}(k)\right)
e_{\sigma\nu}\gamma^\sigma e_{\alpha\mu}\gamma^\alpha\nonumber\\
&&\mbox{} -
q_\tau e_{\beta\mu}\tilde g^{\beta\tau}(k)
e_{\sigma\nu}\gamma^\sigma\lslash{q} -
q_\beta e_{\tau\nu}\tilde g^{\beta\tau}(k)
\lslash{q}e_{\alpha\mu}\gamma^\alpha\,.\nonumber\\
\eeqa
Using the relations in \Eq{qdotge}, it follows that the last two terms
in the above equation vanish, while the remaining terms can be
expressed in the form
\beqa
\tilde g^{\beta\tau}(k)(E_{\nu\sigma\tau}\gamma^\sigma)
(E_{\mu\alpha\beta}\gamma^\alpha) & = &
q^2 e_{\mu\alpha} {e_\nu}^\alpha +
\left(q_\beta q_\tau \tilde g^{\beta\tau}\right)
e_{\mu\alpha} {e_\nu}^\alpha +
i\left(q_\beta q_\tau \tilde g^{\beta\tau}\right)
e_{\mu\alpha} e_{\nu\sigma}\sigma^{\alpha\sigma} \,.
\eeqa
Remembering that
\beq
q_\beta q_\tau \tilde g^{\beta\tau} = q^2
\left[1 - \frac{q^2}{4m^2_\lambda}\right] = \frac{q^2 Q^2}{4m^2_\lambda}\,,
\eeq
the term in the square bracket in \Eq{K2sandwiched} becomes
\beq
\left[
\tilde g^{\beta\tau}(k)(E_{\nu\sigma\tau}\gamma^\sigma)
(E_{\mu\alpha\beta}\gamma^\alpha) -
\left(\frac{q^2}{2m_\lambda}\right)^2 e_{\mu\alpha} {e_\nu}^\alpha\right] =
q_\beta q_\tau \tilde g^{\beta\tau}\left[2 e_{\mu\alpha} {e_\nu}^\alpha +
ie_{\mu\alpha} e_{\nu\beta}\sigma^{\alpha\beta}\right]\,,
\eeq
and therefore,
\beq
\label{K2final}
(\lslash{k} + m_\lambda)K_{\mu\nu 2}(\lslash{k}^\prime - m_\lambda) =
-(\lslash{k} + m_\lambda)\frac{q^2 Q^2}{12 m^2_\lambda}
\left[2 e_{\mu\alpha} {e_\nu}^\alpha +
ie_{\mu\alpha} e_{\nu\beta}\sigma^{\alpha\beta}\right]
(\lslash{k}^\prime - m_\lambda)\,.
\eeq
$K_{\mu\nu 3}$ can be reduced in similar fashion and in that case obtain
\beq
\label{K3final}
(\lslash{k} + m_\lambda)K_{\mu\nu 3}(\lslash{k}^\prime - m_\lambda) =
- (\lslash{k} + m_\lambda)\frac{q^2 Q^2}{12 m^2_\lambda}
\left[2 e_{\mu\alpha} {e_\nu}^\alpha -
ie_{\mu\alpha} e_{\nu\beta}\sigma^{\alpha\beta}\right]
(\lslash{k}^\prime - m_\lambda)\,.
\eeq
Therefore, from \Eqs{K2final}{K3final},
\beq
\label{K23final}
(\lslash{k} + m_\lambda)(K_{\mu\nu 2} + K_{\mu\nu 3})
(\lslash{k}^\prime - m_\lambda) = 
-\frac{q^2 Q^2}{3m^2_\lambda}
e_{\mu\alpha} {e_\nu}^\alpha
(\lslash{k} + m_\lambda)(\lslash{k}^\prime - m_\lambda)\,,
\eeq
and using \Eq{tracescalar},
\beq
\frac{1}{4}\mbox{Tr}\,(\lslash{k} + m_\lambda)(K_{\mu\nu 2} + K_{\mu\nu 3})
(\lslash{k}^\prime - m_\lambda) = \frac{Q^2}{3}\frac{q^2 Q^2}{2m^2_\lambda}
e_{\mu\alpha} {e_\nu}^\alpha\,.
\eeq

For $K_{\mu\nu 4}$ we note that
\beq
\label{K4prelim1}
\mbox{Tr}\, (\lslash{k} + m_\lambda) K_{\mu\nu 4}
(\lslash{k}^\prime - m_\lambda) = \frac{1}{9}\mbox{Tr}\,
(\lslash{k} - m_\lambda) G_\nu (\lslash{k^\prime} + m_\lambda) F_\mu\,,
\eeq
where
\beqa
F_\mu & = & E_{\mu\alpha\beta}
\left(\gamma^\alpha - \frac{k^{\prime\,\alpha}}{m_\lambda}\right)
\left(\gamma^\beta + \frac{k^\beta}{m_\lambda}\right)\nonumber\\
G_\nu & = & E_{\nu\sigma\tau}
\left(\gamma^\tau + \frac{k^\tau}{m_\lambda}\right)
\left(\gamma^\sigma - \frac{k^{\prime\,\sigma}}{m_\lambda}\right)\,,
\eeqa
which using \Eq{defE} and the relations in \Eqs{qdote}{kErel}
can be written in the form
\beqa
F_\mu & = & e_{\alpha\mu}(\gamma^\alpha\lslash{q} - \lslash{q}\gamma^\alpha)
+ \frac{q^2}{m_\lambda}e_{\alpha\mu}\gamma^\alpha \nonumber\\
G_\mu & = & -e_{\alpha\mu}(\gamma^\alpha\lslash{q} - \lslash{q}\gamma^\alpha)
+ \frac{q^2}{m_\lambda}e_{\alpha\mu}\gamma^\alpha \,.
\eeqa
With these forms it is straightforward to obtain
\beqa
(\lslash{k}^\prime + m_\lambda)F_\mu(\lslash{k} - m_\lambda) & = & 
\left(\frac{q^2}{m_\lambda} - 4 m_\lambda\right)
(\lslash{k}^\prime + m_\lambda)
e_{\alpha\mu}\gamma^\alpha
(\lslash{k} - m_\lambda)\nonumber\\
(\lslash{k} - m_\lambda)G_\mu(\lslash{k}^\prime + m_\lambda) & = & 
\left(\frac{q^2}{m_\lambda} - 4 m_\lambda\right)
(\lslash{k} - m_\lambda)
e_{\alpha\mu}\gamma^\alpha
(\lslash{k}^\prime + m_\lambda)\,,
\eeqa
and then using \Eqs{egammasandwiched1}{egammasandwiched2} and the cyclic
property of the trace operation in \Eq{K4prelim1},
\beq
\mbox{Tr}\,(\lslash{k} + m_\lambda)K_{\mu\nu 4}
(\lslash{k}^\prime - m_\lambda) =
-\left(\frac{Q^2}{3m_\lambda}\right)^2
\mbox{Tr}\,(\lslash{k} - m_\lambda)iq^2\tilde\gamma_\nu\gamma_5
(\lslash{k}^\prime + m_\lambda)iq^2\tilde\gamma_\mu\gamma_5 \,,
\eeq
where we have used \Eq{Q2ann}. Therefore,
\beq
\mbox{Tr}\,(\lslash{k} + m_\lambda)K_{\mu\nu 4}
(\lslash{k}^\prime - m_\lambda) =
\left(\frac{Q^2 q^2}{3m_\lambda}\right)^2
\mbox{Tr}\,(\lslash{k} + m_\lambda)\gamma_\nu
(\lslash{k}^\prime + m_\lambda)\gamma_\mu + O(q_\mu,q_\nu)\,,\nonumber\\
\eeq
and finally taking the trace and using \Eq{kkprimetoQmunu}
\beq
\label{K4final}
\mbox{Tr}\,(\lslash{k} + m_\lambda)K_{\mu\nu 4}
(\lslash{k}^\prime - m_\lambda) =
-2\left(\frac{Q^2 q^2}{3m_\lambda}\right)^2 Q_{\mu\nu} + O(q_\mu,q_\nu)\,,
\eeq
where $Q_{\mu]nu}$ is defined in \Eq{Qmunu} and $O(q_\mu,q_\nu)$
stands for terms that are proportional to $q_\mu$ and/or $q_\nu$.

Thus, collecting the formulas given in Eqs.\ (\ref{K1final}), (\ref{K23final}),
and (\ref{K4final}), using \Eq{tracescalar}
and substituting \Eq{ee} for $e_{\alpha\mu} {e^\alpha}_\nu$, we obtain
\beqa
\frac{1}{4}\mbox{Tr}\,(\lslash{k} + m_\lambda)K_{\mu\nu 1}
(\lslash{k}^\prime - m_\lambda) & = &
-\frac{1}{4}\left(\frac{Q^2 q^2}{m_\lambda}\right)^2 Q_{\mu\nu} +
O(q_\mu,q_\nu)\,,\nonumber\\
\frac{1}{4}\mbox{Tr}\,(\lslash{k} + m_\lambda)
(K_{\mu\nu 2} + K_{\mu\nu 3})
(\lslash{k}^\prime - m_\lambda) & = & 
\frac{1}{6}\left(\frac{Q^2 q^2}{m_\lambda}\right)^2 Q_{\mu\nu} +
O(q_\mu,q_\nu)\,,\nonumber\\
\frac{1}{4}\mbox{Tr}\,(\lslash{k} + m_\lambda)K_{\mu\nu 4}
(\lslash{k}^\prime - m_\lambda) & = &
-\frac{1}{18}\left(\frac{Q^2 q^2}{m_\lambda}\right)^2 Q_{\mu\nu} +
O(q_\mu,q_\nu)\,,
\eeqa
which in turn yields
\beq
L^\prime_{\mu\nu} = -\frac{5}{36}\left(\frac{Q^2 q^2}{m_\lambda}\right)^2
Q_{\mu\nu} + O(q_\mu,q_\nu)\,.
\eeq

\section{The $L^{\prime\prime}_{\mu\nu}$ trace}
\label{appendixC}

We recall the following identity that holds for for the trace of
any set of Dirac matrices $A_1, ..., A_n$,
\beq
\mbox{Tr}\,A_1 ... A_n = \mbox{Tr}\,A^c_n ... A^c_1\,,
\eeq
where
\beq
A^c_i \equiv C^{-1}A^T C\,,
\eeq
with $C$ being the $C$ matrix transformation that is defined by
the relation
\beq
C^{-1}\gamma^T C = -\gamma_\mu \,.
\eeq
Using this, the term with the $\gamma_\nu$ in \Eq{Ldprimemunu} can
then be rewritten in the form
\beqa
L^{\prime\prime}_{\mu\nu} & = & \left(\frac{i}{4}\right)
g_{\sigma\tau} E_{\nu\alpha\beta}
\mbox{Tr}\,\gamma_5\gamma_\mu
(\lslash{k} + m_\lambda) R^{\tau\beta}(k) (\lslash{k}^\prime - m_\lambda)
R^{\alpha\sigma}(k^\prime)\nonumber\\
&&\mbox{} -
\left(\frac{i}{4}\right)
g_{\sigma\tau} E_{\mu\alpha\beta}
\mbox{Tr}\,\gamma_5\gamma_\nu
(\lslash{k} - m_\lambda) R^{\tau\beta}(k) (\lslash{k}^\prime + m_\lambda)
R^{\alpha\sigma}(k^\prime)\,.
\eeqa
and with help of \Eq{Ridentities}, we can then write
\beq
\label{LintL1L2}
L^{(int)}_{\mu\nu} = \left(\frac{i}{4}\right)
\left(L^{(1)}_{\mu\nu} - L^{(2)}_{\mu\nu}\right)\,,
\eeq
where
\beqa
\label{L12def}
L^{(1)}_{\mu\nu} & = & E_{\nu\alpha\beta}
\mbox{Tr}\,\gamma_5\gamma_\mu (\lslash{k} + m_\lambda)
J^{\beta\alpha}(k,k^\prime)
(\lslash{k}^\prime - m_\lambda)\nonumber\\
L^{(2)}_{\mu\nu} & = & E_{\mu\alpha\beta}
\mbox{Tr}\,\gamma_5\gamma_\nu (\lslash{k} - m_\lambda)
J^{\beta\alpha}(-k,-k^\prime)
(\lslash{k}^\prime + m_\lambda)\,,
\eeqa
with
\beqa
\label{Jdef}
J^{\beta\alpha} & = & g_{\sigma\tau}
\left[
\tilde g^{\tau\beta}(k) -
\frac{1}{3}\left(\gamma^\tau - \frac{k^\tau}{m_\lambda}\right)
\left(\gamma^\beta + \frac{k^\beta}{m_\lambda}\right)
\right] \nonumber\\
&& \mbox{}\times
\left[\tilde g^{\alpha\sigma}(k^\prime) -
\frac{1}{3}\left(\gamma^\alpha - \frac{k^{\prime\,\alpha}}{m_\lambda}\right)
\left(\gamma^\sigma + \frac{k^{\prime\,\sigma}}{m_\lambda}\right)\right]\,.
\eeqa
This form makes it clear that once we evaluate $L^{(1)}_{\mu\nu}$, then
we can obtain by writing
\beq
\label{L12rel}
L^{(2)}_{\mu\nu} = \left. L^{(1)}_{\nu\mu}
\right|_{m_\lambda\rightarrow -m_\lambda} \,.
\eeq

We begin by evaluating the trace
\beq
\mbox{Tr}\,\gamma_5 \gamma^\mu (\lslash{k} + m_\lambda)
q_\beta J^{\beta\alpha} (\lslash{k}^\prime - m_\lambda)
\eeq
Expanding out \Eq{Jdef} we have
\beqa
q_\beta J^{\beta\alpha} & = & \left[
\tilde g_{\sigma\beta}(k) k^{\prime\,\beta} -
\frac{1}{3}\left(\gamma_\sigma - \frac{k_\sigma}{m_\lambda}\right)
\left(\lslash{k}^\prime + \frac{k\cdot k^\prime}{m_\lambda}\right)\right]
\left[\tilde g^{\alpha\sigma}(k^\prime) -
\frac{1}{3}\left(\gamma^\alpha - \frac{k^{\prime\,\alpha}}{m_\lambda}\right)
\left(\gamma^\sigma + \frac{k^{\prime\,\sigma}}{m_\lambda}\right)\right]
\nonumber\\
& \equiv &
\left(J^{\alpha}\right)_{11} +
\left(J^{\alpha}\right)_{12} +
\left(J^{\alpha}\right)_{21} +
\left(J^{\alpha}\right)_{22}
\eeqa
where
\beqa
\left(J^{\alpha}\right)_{11} & = &
\tilde g_{\sigma\beta}(k) k^{\prime\,\beta}
\tilde g^{\alpha\sigma}(k^\prime)\nonumber\\
\left(J^{\alpha}\right)_{12} & = & -\frac{1}{3}
\tilde g_{\sigma\beta}(k) k^{\prime\,\beta}
\left(\gamma^\alpha - \frac{k^{\prime\,\alpha}}{m_\lambda}\right)
\left(\gamma^\sigma + \frac{k^{\prime\,\sigma}}{m_\lambda}\right)\nonumber\\
\left(J^{\alpha}\right)_{21} & = & -\frac{1}{3}
\tilde g^{\alpha\sigma}(k^\prime)
\left(\gamma_\sigma - \frac{k_\sigma}{m_\lambda}\right)
\left(\lslash{k}^\prime + \frac{k\cdot k^\prime}{m_\lambda}\right)\nonumber\\
\left(J^{\alpha}\right)_{22} & = & \frac{1}{9}
\left(\gamma_\sigma - \frac{k_\sigma}{m_\lambda}\right)
\left(\lslash{k}^\prime + \frac{k\cdot k^\prime}{m_\lambda}\right)
\left(\gamma^\alpha - \frac{k^{\prime\,\alpha}}{m_\lambda}\right)
\left(\gamma^\sigma + \frac{k^{\prime\,\sigma}}{m_\lambda}\right)
\eeqa 
Then we get
\beqa
\mbox{Tr}\,\gamma_5 \gamma^\mu (\lslash{k} + m_\lambda)
\left(J^{\alpha}\right)_{ab}
(\lslash{k}^\prime - m_\lambda) & = &
C_{ab}\mbox{Tr}\,\gamma_5 \gamma_\mu (\lslash{k} + m_\lambda)\gamma^\alpha
(\lslash{k}^\prime - m_\lambda)\nonumber\\
\eeqa
where
\beqa
C_{11} & = & 0 \nonumber\\
C_{12} & = &
-\frac{1}{3}\left(\frac{k\cdot k^\prime}{m_\lambda}\right)
\left(1 - \frac{k\cdot k^\prime}{m^2_\lambda}\right)\nonumber\\
C_{21} & = & \frac{1}{3} m_\lambda
\left(1 - \frac{k\cdot k^\prime}{m^2_\lambda}\right)\nonumber\\
C_{22} & = & 
\frac{1}{9}\left(\frac{k\cdot k^\prime}{m_\lambda}\right)
\left(1 - \frac{k\cdot k^\prime}{m^2_\lambda}\right)
\eeqa
Therefore
\beqa
\label{trace1}
\mbox{Tr}\,\gamma_5 \gamma^\mu (\lslash{k} + m_\lambda)
q_\beta J^{\beta\alpha} (\lslash{k}^\prime - m_\lambda) & = &
C\mbox{Tr}\,\gamma_5 \gamma_\mu (\lslash{k} + m_\lambda)\gamma^\alpha
(\lslash{k}^\prime - m_\lambda) \nonumber\\
& = & C\mbox{Tr}\,\gamma_5\gamma^\mu\lslash{k}\gamma^\alpha\lslash{k}^\prime
\nonumber\\
& = & 4iC\epsilon^{\mu\alpha\lambda\rho} k_\lambda k^\prime_\rho \nonumber\\
& = & 2iC e_{\mu\alpha}\,,
\eeqa
where
\beq
C = \left(\frac{k\cdot k^\prime}{m^2_\lambda} - 1\right)
\left(\frac{2}{9}\frac{k\cdot k^\prime}{m_\lambda} -
\frac{1}{3}m_\lambda\right) =
\frac{Q^2 (Q^2 + m^2_\lambda)}{18 m^3_\lambda} \,.
\eeq
Similarly,
\beqa
\label{trace2}
\mbox{Tr}\,\gamma_5 \gamma^\mu (\lslash{k} + m_\lambda)
q_\alpha J^{\beta\alpha} (\lslash{k}^\prime - m_\lambda) & = &
-C\mbox{Tr}\,\gamma_5 \gamma_\mu (\lslash{k} + m_\lambda)\gamma^\alpha
(\lslash{k}^\prime - m_\lambda) \nonumber\\
& = & -C\mbox{Tr}\,\gamma_5\gamma^\mu\lslash{k}\gamma^\beta\lslash{k}^\prime
\nonumber\\
& = & -4iC\epsilon^{\mu\beta\lambda\rho} k_\lambda k^\prime_\rho\nonumber\\
& = & -2iC e_{\mu\beta}\,.
\eeqa
Using \Eq{trace1} in \Eq{L12def} we then obtain,
\beq
L^{(1)}_{\mu\nu} = -4iC e_{\alpha\mu} {e^{\alpha}}_\nu \,,
\eeq
while \Eq{L12rel}, yields
\beq
L^{(2)}_{\mu\nu} = - L^{(1)}_{\mu\nu} \,.
\eeq
From \Eq{LintL1L2} we then obtain finally
\beq
L^{\prime\prime}_{\mu\nu} = \frac{Q^2 (Q^2 + m^2_\lambda)}{9 m^3_\lambda}
e_{\alpha\mu} {e^{\alpha}}_\nu \,,
\eeq
where $e_{\alpha\mu} {e^{\alpha}}_\nu$ is given explicitly in \Eq{ee}.

\end{document}